\documentclass[aps,pra,twocolumn,nofootinbib, superscriptaddress,10pt]{revtex4-1}

\usepackage{amssymb,mathtools}
\usepackage{amsthm}
\usepackage{multirow}
\usepackage{graphicx}

\usepackage{bm}
\usepackage{color}
\usepackage{caption}

\usepackage[caption = false]{subfig}

\DeclareCaptionLabelFormat{cont}{#1~#2\alph{ContinuedFloat}}
\usepackage{mleftright}
\usepackage{framed}
\usepackage[framemethod=TikZ]{mdframed}
\usepackage{tikz}
\usetikzlibrary{tikzmark}
\usetikzlibrary{shapes,snakes}
\usetikzlibrary{fit,backgrounds}
\usetikzlibrary{decorations.pathreplacing,angles,quotes}
\usetikzlibrary{patterns}
\usetikzlibrary{calc}
\usetikzlibrary{positioning}
\tikzstyle{block} = [rectangle, draw, 
    minimum width=4em, text centered, rounded corners, minimum height=1em]
\tikzstyle{network} = [rectangle, draw, 
    minimum width=10em, text centered, rounded corners, minimum height=6em]
\tikzstyle{rec} = [rectangle, draw]
\tikzstyle{line} = [draw, -latex, thick]

\newcommand*{\QEDA}{\hfill\ensuremath{\blacksquare}}%

\newtheorem{lemm}{Lemma}

\newtheorem{theo}{Theorem}

\def\FF{\mathbb{F}}

\DeclarePairedDelimiter\bra{\langle}{\rvert}
\DeclarePairedDelimiter\ket{\lvert}{\rangle}
\DeclareMathOperator{\rank}{rank}

\def\cH{\mathcal{H}}

\def\bZ{\mathbb{Z}}
\def\sX{\mathsf{X}}
\def\sZ{\mathsf{Z}}
\def\bX{\mathbf{X}}
\def\bZ{\mathbf{Z}}
\def\bW{\mathbf{W}}
\def\bs{\mathbf{s}}
\def\bt{\mathbf{t}}
\def\bx{\mathbf{x}}
\def\by{\mathbf{y}}

\def\Tr{\mathop{\rm Tr}\nolimits}
\def\tr{\mathop{\rm tr}\nolimits}
\def\Ker{\mathop{\rm Ker}\nolimits}

\begin{document}
\title{Quantum Capacity of Partially Corrupted Quantum Network}

\author{Masahito Hayashi}
\email{masahito@math.nagoya-u.ac.jp}
\affiliation{Graduate School of Mathematics, Nagoya University, Nagoya, 464-8602, Japan}

\affiliation{Shenzhen Institute for Quantum Science and Engineering,
	Southern University of Science and Technology,
	Shenzhen,
	518055, China}
\affiliation{Center for Quantum Computing, Peng Cheng Laboratory, Shenzhen 518000, China}

\affiliation{Centre for Quantum Technologies, National University of Singapore, 3 Science Drive 2, 117542, Singapore}

\author{Seunghoan Song}
\email{m17021a@math.nagoya-u.ac.jp}

\affiliation{Graduate School of Mathematics, Nagoya University, Nagoya, 464-8602, Japan}

\begin{abstract}
We discuss a quantum network, in which 
the sender has $m_0$ outgoing channels,
the receiver has $m_0$ incoming channels, 
each channel is of capacity $d$,
each intermediate node applies invertible unitary,
only $m_1$ channels are corrupted, 
and other non-corrupted  channels are noiseless.
As our result, we show that the quantum capacity is not smaller than 
$(m_0-2m_1+1)\log d$ 
under the following two settings.
In the first case, the unitaries on intermediate nodes are arbitrary and
the corruptions on the $m_1$ channels are individual.
In the second case, the unitaries on intermediate nodes are restricted to Clifford operations and
the corruptions on the $m_1$ channels are adaptive, i.e.,
the attacker is allowed to have a quantum memory.
Further, 
our code in the second case
realizes the noiseless communication even with the single-shot setting
and is constructed dependently only on 
the network topology and the places of the $m_1$ corrupted channels
while this result holds regardless of the network topology and the places.
\end{abstract}

\date{\today}
\maketitle


\section{Introduction}%
When two distant players communicate their quantum states via a single channel, 
their communication can be disturbed by the corruption of the channel.
In the classical information theory,
to resolve this problem, they employ information transmission over a network, which is composed of nodes and channels \cite{Cai06,Jaggi2008,Yao2014}.
Combining a network code, they realized a reliable communication 
even when a part of nodes are corrupted
because a network code realizes the diversification of risk.
Many existing papers for the quantum network addressed the
multiple-unicast network \cite{Hayashi2007,PhysRevA.76.040301,Kobayashi2009,Leung2010,Kobayashi2010,Kobayashi2011,JFM11,OKH17,OKH17-2,SH18-1}.
The coding scheme proposed in \cite{PhysRevA.76.040301} was already implemented experimentally \cite{Lu}.
However, they did not discuss a reliable quantum network code over 
corruptions on the quantum network even in the unicast network.
Since a corruption on a node can be propagated to the entire network
due to the network structure,
it is desired to discuss a reliable quantum network code 
for the unicast network even in the presence of the corruption.
Indeed, while the preceding paper \cite{SH18-2} constructed such a network code, it did not discuss the optimality of the transmission rate.
Also, the code constructed in \cite{SH18-2} works only when 
the unitary operations on the network nodes belong to a limited class.

This paper discusses the quantum capacity of a partially corrupted quantum unicast network, i.e., the optimal value of the reliable transmission rate under the knowledge of the form of the corruption
for a more general class of node operations on the network.
Our problem setting is slightly different from those of the preceding papers \cite{SH18-1,SH18-2} because \cite{SH18-1,SH18-2} discussed quantum network coding 
under limited knowledge of the corruption.
Whereas conventional network coding considers the optimization of node operations given a directed graph of the network, 
we consider the quantum capacity when 
node operations are given as well 
because it is often quite difficult to control node operations.
Specifically, we address the worst-case capacity among a certain condition, which is formulated as follows.

Every quantum channel transmits a $d$-dimensional system by one use of the network.
The sender has $m_0$ outgoing quantum channels.
Each intermediate node has the same number of incoming quantum channels and outgoing quantum channels.
The node applies a fixed unitary across the incoming quantum systems and outputs them to the outgoing quantum channels.
Finally, the receiver receives $m_0$ quantum systems via $m_0$ incoming quantum channels.
We assume that only $m_1$ quantum channels are corrupted at most.
Other channels are assumed to be noiseless
in the same way as secure classical network coding \cite{Cai06,Jaggi2008,Yao2014}
because the errors of these normal channels can be corrected by quantum error correcting code.
Also, the network is assumed to have no cycle
and to be well synchronized, i.e., to have no delayed transmission.
Only the sender and the receiver are allowed to optimize their coding operation
due to the difficulty of node operation control.
Although the paper \cite{SH18-2} assumed that the sender and the receiver do not know the places of 
corrupted channels,
the places are assumed to be known to them in our setting.

 There are two types of quantum settings.
The first one is the individual corruption, in which
the corruption on each corrupted quantum channel is done individually.
The other is the adaptive corruption, in which
the corruptions on respective corrupted quantum channels are done adaptively.
That is, the attacker has a quantum memory, and 
the quantum memory interacts the corrupted quantum channel on each corruption.
For an adversarial corruption, we need to consider such a malicious case.
This kind of corruption can be written by quantum comb \cite{PhysRevLett.101.060401,PhysRevA.80.022339}.
The adaptive setting is more general than the individual setting,
and the adaptive setting often cannot be reduced to the individual setting
in general.
For example, adaptive strategies cannot be reduced to individual strategies 
quantum channel discrimination \cite{HHLW}.

In the classical case of this setting,
we can show that the capacity, i.e., the maximum transmission rate is not smaller than $(m_0-m_1)\log d$.
In contrast, when our quantum channel has only individual corruptions,
we find that the quantum capacity, i.e., the maximum transmission rate of the quantum state 
is not smaller than $(m_0-2 m_1+1)\log d$.
This fact is shown by the analysis of coherent information on the quantum network.
Further, when the unitaries on our network are limited to Clifford operations,
the quantum capacity is not smaller than $(m_0-2 m_1+1)\log d$
even when the corruptions are adaptive.
In this case, our code can be constructed in the single-shot setting by using Clifford operations.
This construction depends only on the applied Clifford operations and the places of the corrupted channels, and
is independent of Eve's operation to the corrupted channels.
This phenomena can be intuitively explained for the case of Clifford operations as follows.
The sender can identify the place of the first corruption.
However, at the other corruptions, the corrupted computation bases and the corrupted Fourier bases split in general.
For this characterization, the behavior of the error is described by the 
the symplectic structure.
In particular, the symplectic diagonalization plays an essential role in the code construction.
Hence, in the worst case, totally $2 m_1-1$ quantum systems are corrupted.

\section{Classical network model}
When all the node operations are invertible linear on a finite field {of order} $q$ ($=d$),
the receiver can find a linear subspace for corrupted information, as discussed in \cite{Jaggi2008}.
The dimension of the subspace is bounded by $m_1$.
Hence, the capacity is not smaller than $(m_0-m_1)\log {d}$.
However, when node operations are not necessarily linear but are invertible,
we cannot apply the above discussion.
Even in this case, 
we can show that the capacity is not smaller than $(m_0-m_1)\log d$ as 
shown in Appendix A.

\begin{figure}[t]
\includegraphics[width=8cm]{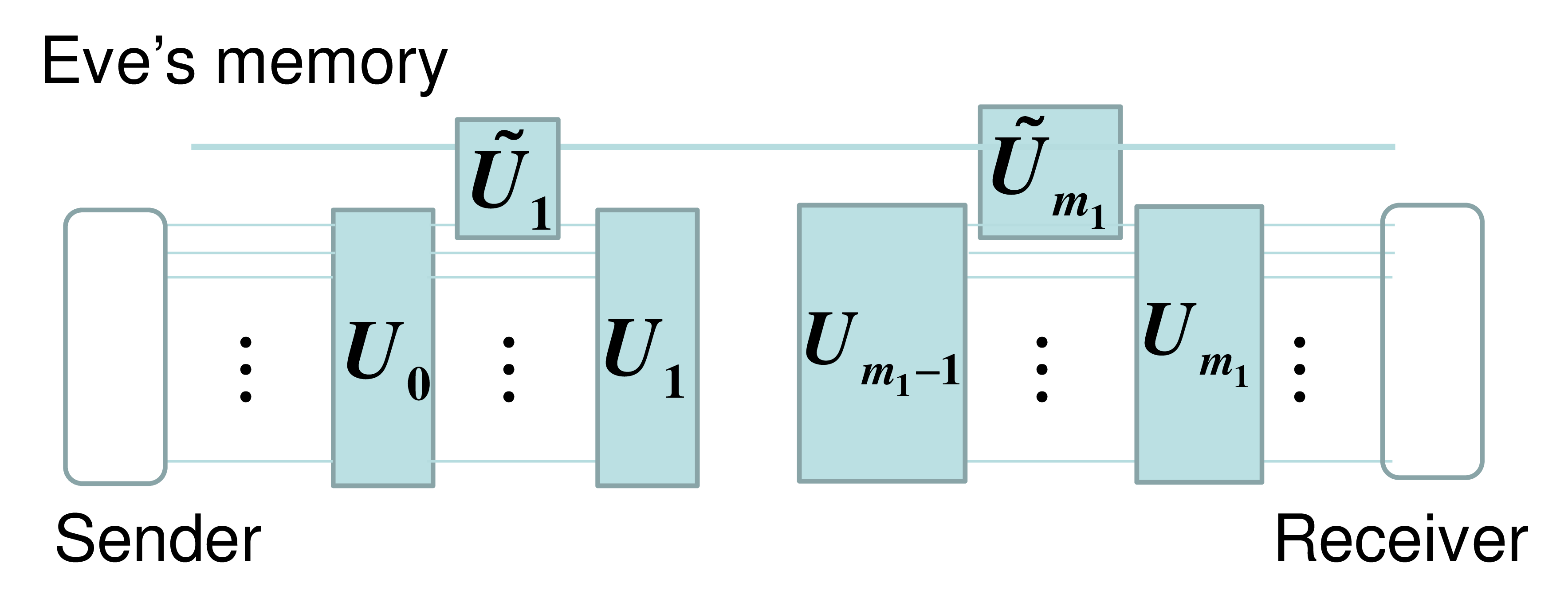}	
\caption{Unitary network with adaptive corruption.
After the application of the unitary $U_{i-1}$,
the unitary $\tilde{U}_i$ is applied.}
\label{Fig1}
\end{figure}%

\section{General unitary network model}
\begin{figure}
\subfloat[Quantum network with three intermediate nodes. 
Sender and receiver have $m_0=6$ outgoing and incoming channels, respectively, and 
each intermediate node has the same number of incoming and outgoing channels.
The zigzagged lines are corrupted channels and two channels are corrupted (i.e., $m_1= 2$).]{
\centering
\begin{tikzpicture}[transform shape, node distance = 5cm, auto, every node/.style={outer sep=0}]
    \node [circle,draw,inner sep = 0.2em] (source) {$v_0$};
    \node [below left=-0.1em and -1em of source] {Sender};
    \node [circle,draw, above right=4em and 4em of source,inner sep = 0.2em] (node1) {$v_1$};
    \node [above=-0.1em of node1] {$V_1$};
    \node [circle,draw, below right=4em and 6em of source,inner sep = 0.2em] (node2) {$v_2$};
    \node [below=-0.1em of node2] {$V_2$};
    \node [circle,draw, right=10em of source,inner sep = 0.2em] (node3) {$v_3$};
    \node [above=-0.1em of node3] {$V_3$};
    \node [circle,draw, right=18em of source,inner sep = 0.2em] (target) {$v_4$};
    \node [below right=0.1em and -1em of target] {Receiver};
    
    \path [line] (source) edge [bend left=30] (target);
    \path [line] (source) edge [bend left=25]  (node1);
    \path [line] (source) edge [bend left=0]  (node1);
    
    \path [line] (node1) edge [bend left=25,->,decorate,decoration={snake,amplitude=.4mm,segment length=2mm,post length=1mm}] node{$\tilde{U}_2$}(target);
    \path [line] (node1) edge (node2);
    
    \path [line] (source) edge [bend left=15]  (node2);
    \path [line] (source) edge [bend right=15]  (node2);
    
    \path [line] (source) edge  (node3);

    \path [line] (node2) edge [bend left=15,->,decorate,decoration={snake,amplitude=.4mm,segment length=2mm,post length=1mm}] node{$\tilde{U}_1$}(node3);
    \path [line] (node2) edge [bend right=15] (node3);
    \path [line] (node2) edge [bend right=15] (target);
    
    \path [line] (node3) edge [bend left=20] (target);
    \path [line] (node3) edge [bend right=20] (target);
    \path [line] (node3) edge  (target);
\end{tikzpicture}} \\
\subfloat[The network in (a) in the form of Fig. \ref{Fig1}.]{
\centering
\includegraphics[width=8cm]{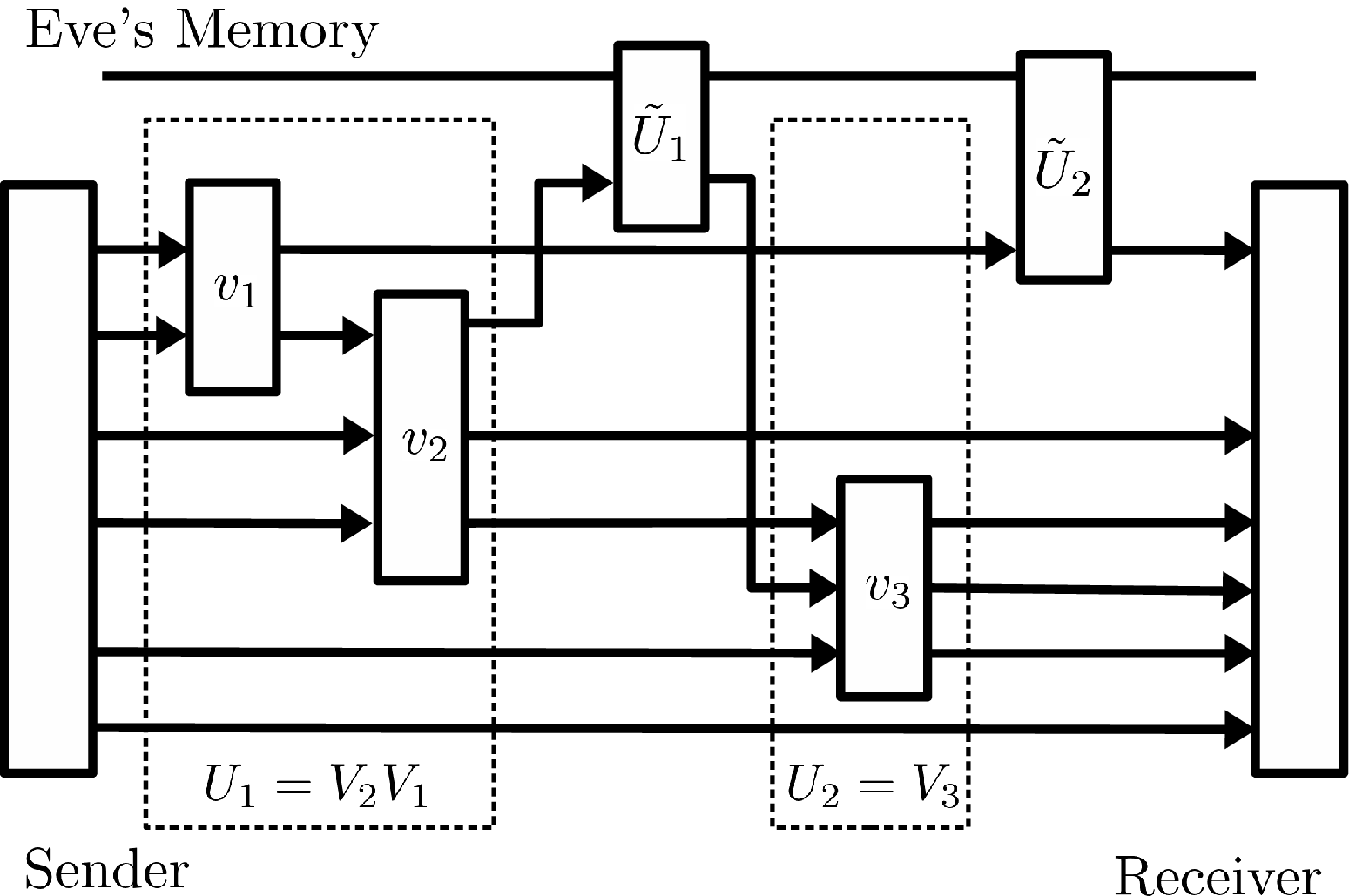}}
\caption{{Reorganization of the network in the form of Fig. \ref{Fig1}.} }   \label{fig:1}
\end{figure}

The general unitary network model is described as follows.
In this model, we assume that the places of the channels to be corrupted are known.
Since our network is composed of unitary operations and partial corruptions,
our network model of the adaptive corruption is given as 
the general form with Fig. \ref{Fig1}, whose reason is illustrated in Fig. \ref{fig:1}.
The input and output systems are the $m_0$-tensor product system ${\cal H}^{\otimes m_0}$ of 
the same system ${\cal H}$ of dimension $d$, and
$m_1+1$ unitaries ${\bm U}=(U_0, U_1,\ldots, U_{m_1})$ are applied between 
the input and output systems, which has $m_1$ intervals.
Eve can access only the first system on each interval,
and has her memory so that 
the corruption in the $i$-th interval is given as the unitary $\tilde{U}_i$ between 
her memory and the corrupted system, i.e., the first system on the $i$-th interval.

\begin{figure}[t]
\includegraphics[width=8cm]{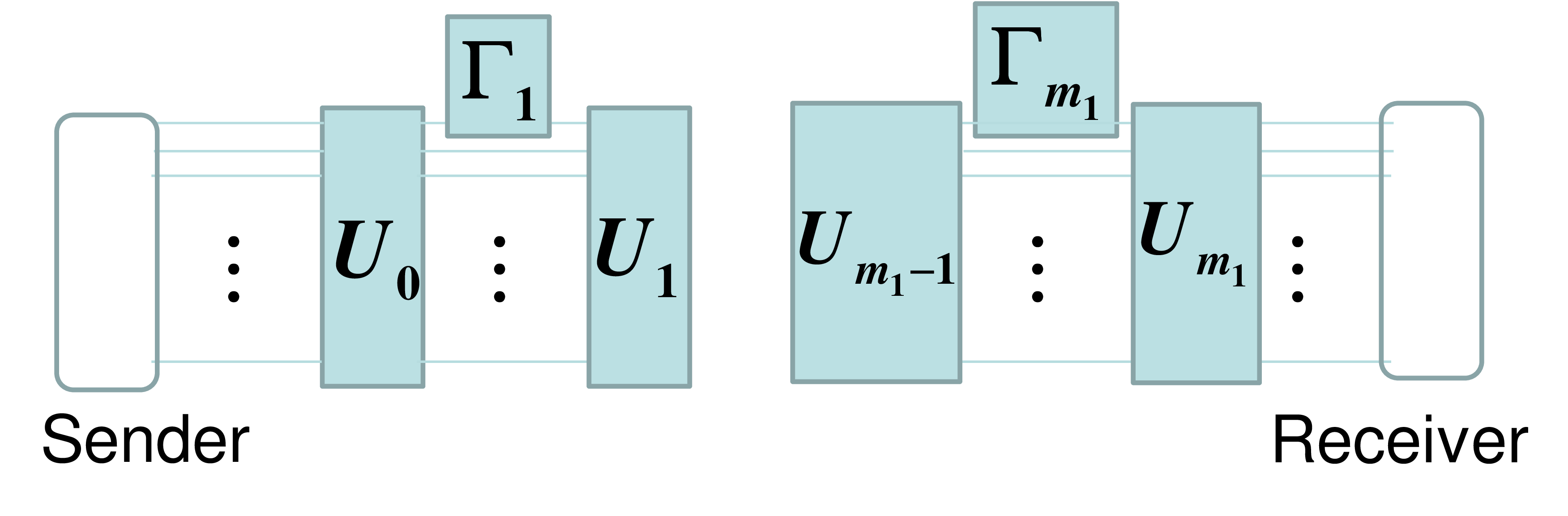}	
\caption{Unitary network with individual corruption.
After the application of the unitary $U_{i-1}$,
the noisy operation $\Gamma_i$ is applied.}
\label{Fig2}
\end{figure}

Our first result is on the minimum capacity of the general unitary network 
with individual corruption, in which Eve is assumed to have no memory.
Hence, her operation on the $i$-th interval
can be written as TP-CP maps ${\bm \Gamma}=
(\Gamma_1, \ldots, \Gamma_{m_1})$ as Fig. \ref{Fig2}.
In this case, the channel between 
the input and output systems is denoted by
$\Lambda({\bm U}, {\bm \Gamma}) $.
$C(\Lambda)$ expresses the quantum capacity of a quantum channel $\Lambda$.
One of our main results characterizes the quantum capacity
under the individual corruption as follows.

\begin{theo}\label{Th1}
The minimum quantum capacity is given as follows.
\begin{align}
\min C(\Lambda({\bm U}, {\bm \Gamma}) )
=
(m_0-2m_1+1)\log d.\label{MS}
\end{align}
Here, the minimum is taken 
over all channels $\Lambda({\bm U}, {\bm \Gamma})$
under the individual corruption.
\end{theo}

To show the theorem, 
we employ the coherent information $I_c(\rho,\Lambda)$ for 
an input state $\rho$ and
a quantum channel $\Lambda$ with an output system $A$.
By using the environment $E$ of $\Lambda$, 
the coherent information $I_c(\rho,\Lambda)$ is written as
$H(A)-H(E)$, where $H(A)$ and $H(E)$ are the von Neumann entropy of the respective system 
when the input state is $\rho$ \cite{SN96}\cite[(8.37)]{Haya2}.
It is known that 
the quantum capacity $C(\Lambda)$ is given as
the maximum 
\begin{align}
\lim_{n \to \infty}\max_{\rho} \frac{1}{n}I_c(\rho,\Lambda^{\otimes n}),
\label{Max-Mutual}
\end{align}
where the maximum is taken over all the input densities on the $n$-tensor system of the input system of $\Lambda$ \cite{Barnum97,Lloyd,Shor,Devetak}, \cite[Theorem 9.10]{Haya2}.
  
\begin{lemm}\label{L3-1}
A individual corruption
$\Lambda({\bm U}, {\bm \Gamma}) $ satisfies

\begin{align}
\max_{\rho}I_c(\rho, \Lambda({\bm U}, {\bm \Gamma}) )
\ge 
(m_0-2m_1+1)\log d.
\end{align}
\end{lemm}
The proof of Lemma \ref{L3-1} is given in Appendix B.
Due to the additivity property $
I_c(\rho^{\otimes n},\Lambda^{\otimes n})=n I_c(\rho,\Lambda)$,
the lemma and the above capacity formula \eqref{Max-Mutual} guarantee the inequality $\ge$
in \eqref{MS}.
As shown later, 
Theorem \ref{Th2} guarantees the existence of a channel 
$\Lambda({\bm U}, {\bm \Gamma})$ to satisfy the equality,
which completes the proof of Theorem \ref{Th1}.

\section{Clifford network model}
To extend Theorem \ref{Th1} to the adaptive corruption described in Fig. \ref{Fig1}, 
we introduce Clifford network model.
In this model, our code can achieve the capacity even with the single-shot setting. 
For this aim, we prepare several notations.
Given a prime power $q=p^{d_q}$, 
our Hilbert space $\cH$ is assumed to be spanned by 
the computational basis $\{\ket{x}\}_{x \in \FF_q}$,
where $\FF_q$ is the algebraic extension of the finite field $\FF_p$
with degree $d_q$.
That is, the dimension of the Hilbert space $\cH$ is assumed to be $q$.
Then, for $s,t \in \FF_q$,
we define the generalized Pauli operators $\sX(s)$ and $\sZ(t)$ 
as $\sX(s):=\sum_{x \in \FF_q}\ket{x+s}\bra{x}$
and $\sZ(t):=\sum_{x \in \FF_q}\omega^{\tr xt}\ket{x}\bra{x}$,
where $\omega:= e^{2\pi\sqrt{-1}/p }$.
Here, for an element $z \in \FF_q$,
$\tr z $ expresses the element $\Tr M_z \in \FF_p$, 
where $M_z$ denotes the matrix representation of the multiplication map $x \mapsto zx $ 
with identifying the finite field $\FF_{q}$ with the vector space $\FF_p^{d_q}$.
We define the Fourier basis 
$\left\{ \ket{y}_{F}\in \cH \right\} _{y \in \FF_q}
$
of the computational basis $\left\{ \ket{x} \right\} _{x \in \FF_q} \in \cH$
as 
\[
\ket{y}_F:=\sum_{x \in \FF_q} \frac{1}{\sqrt{q}}\omega^{\tr xy }\ket{x}.
\]

To consider our network model,
for vectors $\bs=(s_1, \ldots, s_n),\bt =(t_1, \ldots, t_n) \in \FF_q^n$,
we define 
the operators 
$\bX(\bs)$ and $\bZ(\bt)$
on the $n$-fold tensor product system $\cH^{\otimes n}$ as
$\bX(\bs) := \sX(s_1)\otimes \cdots \otimes \sX(s_n)$ and 
$\bZ(\bt) := \sZ(t_1)\otimes \cdots \otimes \sZ(t_n)$.
Then, the discrete Weyl operator is defined as
$\bW(\bs,\bt):= \bX(\bs)\bZ(\bt)$.
Then, for $(\bs,\bt),(\bs',\bt')\in \FF_q^{2n}$,
we define 
the skew symmetric matrix $J$ on $\FF_q^{2n}$
and
the inner product $\langle (\bs,\bt),(\bs',\bt')\rangle \in \FF_p$
as
$J=
\left(
\begin{array}{cc}
0 & -I \\
I & 0
\end{array}
\right)$
with $\langle (\bs,\bt),(\bs',\bt')\rangle:=
\sum_{i=1}\tr (s_i s_i'+t_i t_i')$.
Then, the commutation relation
\begin{align}
\bW(\bs,\bt)\bW(\bs',\bt')
=
\omega^{\langle (\bs,\bt),J (\bs',\bt')\rangle}
\bW(\bs',\bt') \bW(\bs,\bt)
\end{align}
holds, and a square matrix $g$ on $\FF_q^{2n}$
is called a symplectic matrix
when 
$\langle (\bs,\bt),J(\bs',\bt')\rangle=
\langle g(\bs,\bt),Jg(\bs',\bt')\rangle $
for $(\bs,\bt),(\bs',\bt')\in \FF_q^{2n}$.

Next, we introduce Clifford group as a subset of the set ${\cal U}(\cH^{\otimes n})$ of unitaries on $\cH^{\otimes n}$.
Using the set ${\cal W}:= \{c \bW(\bs,\bt)\}_{|c|=1, (\bs,\bt)\in \FF_q^{2n}}$,
we define the Clifford group ${\cal C}$ as
${\cal C}:= \{U \in {\cal U}(\cH^{\otimes n})|
U {\cal W} U^{-1}={\cal W} \}$.
An element of ${\cal C}$ is called a Clifford unitary.

For any element $U \in {\cal C}$,
there exists a symplectic matrix $g$ such that
\begin{align}
U \bW(\bs,\bt)U^{-1}= c \bW(g(\bs,\bt))\label{MN1}
\end{align}
for $(\bs,\bt)\in \FF_q^{2n}$ with a complex number $c$ satisfying $|c|=1$.
Conversely, 
for any symplectic matrix $g$,
there exists a unitary $U$ to satisfy \eqref{MN1}.
A typical construction of such a unitary $U$ is given in \cite[Section 8.3]{Haya2}.
This construction is called metaplectic representation
and is denoted by $U(g)$ in this paper.

Now, the input and output systems are assumed to be $\cH^{\otimes m_0}$,
and the unitary $U_i$ is to be an element of Clifford group.
Such a network is called Clifford network.
We choose a symplectic matrix $g_i$ as
$U(g_i)=U_i$.

\begin{theo}\label{Th2}
For Clifford network, the minimum quantum capacity is 
$(m_0-2m_1+1)\log q$
in the adaptive corruption, i.e., the case 
when Eve has a memory to perform adaptive attacks.

\end{theo}

To show Theorem \ref{Th2},
we describe the behaviors of the errors 
in the terms of the symplectic structure.
That is, the errors can be described by vectors in $\FF_q^{2m_0}$.
For this aim,
we introduce notations and parameters of the network as follows.
Let $e_i$ be the vector in $\FF_q^{2m_0}$ that has only one nonzero element $1$ in the $i$-th entry.
Using 
$e_1$ 
and
$e_{m_0+1}$, 
we define 
$2m_1$ vectors $v_1, \ldots, v_{2m_1} \in \FF_q^{2m_0}$
as
$v_i:= g_{0}^{-1}\cdots g_{i-1}^{-1} e_1$ 
and
$v_{m_1+i}:= g_{0}^{-1}\cdots g_{i-1}^{-1} e_{m_0+1}$ for $i=1, \ldots, m_1$.
Since $e_1$ and $e_{m_0+1}$ describe the directions of errors in the respective
interval,  
all the directions in the linear space ${\cal V}$ spanned by $v_1, \ldots, v_{2m_1}$
are corrupted in this whole network.
However, since the direction $v_i$ is corrupted, 
the direction $J v_i$ is also corrupted.
Therefore, the set of all corrupted directions is given by ${\cal V}+J{\cal V}$.

In this paper, when a matrix $P$ satisfies $P^2=P$ and ${\rm Im}P={\cal V}$, it is called a projection onto ${\cal V}$.
Then, we choose a projection $P_{{\cal V}}$ onto ${\cal V}$.
Since $P_{{\cal V}}^{\top} J P_{{\cal V}}$ is also an anti-symmetric matrix,
the rank of $P_{{\cal V}}^{\top} J P_{{\cal V}}$ is an even number.
The rank of the matrix $P_{{\cal V}}^{\top} J P_{{\cal V}}$ equals
the rank of matrix $( \langle v_i, J v_j\rangle)_{i,j}$.
Hence, the rank of $P_{{\cal V}}^{\top} J P_{{\cal V}}$ does not depend on the choice of the projection $P_{{\cal V}}$ onto ${\cal V}$
while the choice of the projection $P_{{\cal V}}$ onto ${\cal V}$
is not unique.
With these observations, we define the integers $m_*$ and $m_{**}$
as $m_*=(\rank P_{{\cal V}}^{\top} J P_{{\cal V}})/2$ and 
$m_{**}=\dim {\cal V}-m_{*}$.

Since the rank of the submatrix 
$( \langle v_i,J v_j\rangle )_{i,j=1, m_0+1}$ is 2, 
the rank of $( \langle v_i,J v_j\rangle )_{i,j=1}^{2m_1}$
is at least 2.
As the rank of $( \langle v_i,J v_j\rangle )_{i,j=1}^{2m_1}$
equals the rank of $P_{{\cal V}}^{\top} J P_{{\cal V}}$, 
the inequality 
\begin{align}
m_{*}\ge 1
\label{G0}
\end{align}
holds.
Thus, the inequality $ \dim {\cal V}\ge \rank P_{{\cal V}}^{\top} J P_{{\cal V}}$
implies 
\begin{align}
m_{**}\ge m_*.
\label{G02}
\end{align}
So, since $2 m_1\ge m_{**}+m_* $, we have
\begin{align}
m_{**} \le  2 m_1-1.
\label{G1}
\end{align}

The quantum capacity $C$ is characterized as follows.

\begin{lemm}\label{L2-1}
The capacity $C$ is lower bounded as $C \ge (m_0-m_{**})\log q$ 
in the adaptive case, i.e., the case 
when Eve has a memory to perform her attack.
\end{lemm}

It is a key point for the construction of a code achieving the rate $(m_0-m_{**})\log q$ 
to avoid the space ${\cal V}+J{\cal V}$
from the encoded space.
As shown in Appendix D,
we can choose $2 m_0$ independent vectors 
$w_1, \ldots, w_{m_0}$ and $w_1', \ldots, w_{m_0}'$ satisfying the following conditions.
(i) 
$\langle w_i',J w_j \rangle =\delta_{i,j}$ and
$\langle w_i,J w_j \rangle =\langle w_i',J w_j' \rangle =0$ 
for $i,j=1, \ldots, m_0$.
(ii)
The space $ {\cal V}+J{\cal V}$ is spanned by 
$w_{m_0-m_{**}}, \ldots, w_{m_0}$ and $w_{m_0-m_{**}}', \ldots, w_{m_0}'$.
The condition (i) enables us to choose a symplectic matrix $g_*$ such that
$g_* e_i=  w_i$ and $g_* e_{m_0+i} =  w_i'$ for $i=1, \ldots , m_0$
because the vectors $w_i$ and $w_i'$ with $i=1, \ldots , m_0$
have the same symplectic structure as the vectors $e_1,\ldots, e_{2m_0}$.

Then, we define the encoding unitary $U_e:= U(g_*) $
and the decoding unitary $U_d:= U(g_*)^{-1} U_0^{-1}\cdots U_{m_1}^{-1} $.
The message space is set to $\cH^{\otimes m_0-m_{**}}$.
The encoder is given as follows.
We fix an arbitrary density $\rho_0$ on $\cH^{\otimes m_{**}}$. 
For any input density $\rho_i$ on $\cH^{\otimes m_0-m_{**}}$,
the encoder is given as $\rho_i \mapsto U_e (\rho_i \otimes \rho_0)U_e^{-1}$.
The decoder is given as $\rho \mapsto \Tr_{m_{**}} (U_d \rho U_d^{-1})$,
where $\Tr_{m_{**}}$ is the partial trace with respect to the system $\cH^{\otimes m_{**}}$.

To analyze this code, we remind the following facts.
First, the set of all corrupted directions is given by ${\cal V}+J{\cal V}$.
Second, 
the information on the computational basis 
and the information on the Fourier basis are decoded perfectly,
if and only if the transmitted quantum state is decoded perfectly \cite{Renes}, \cite[Section 8.15]{Haya2}.
It is clear from the first fact and the construction of $U_e$ and $U_d$ that 
the information on the computation basis and the information on the Fourier basis is decoded perfectly
even when Eve makes an adaptive corruption.
Therefore, even for any adaptive corruption,
the pair of the above encoder and the above decoder
can decode the original state $\rho_i$ on $\cH^{\otimes m_0-m_*}$.
This discussion shows Lemma \ref{L2-1}.
Our code construction depends only on 
$g_0, \ldots, g_{m_1}$. That is, it is independent of the remaining unitaries $\tilde{U}_1, \ldots, \tilde{U}_{m_1}$ of Eve's corruption.
The tightness of the evaluation in Lemma \ref{L2-1} is guaranteed as follows.

\begin{lemm}\label{L2-2}
When Eve changes the state on the corrupted edge to the completely mixed state,
the capacity $C$ equals $(m_0-m_{**})\log q$.
\end{lemm}
Lemma \ref{L2-2} is shown in Appendix C.
In this way, the capacity is characterized by $\rank P_{{\cal V}}^{\top} J P_{{\cal V}}$
and $\dim {\cal V}$.
The following lemma clarifies the possible range of these two numbers.

\begin{lemm}\label{L6}
The following conditions are equivalent for two integers $l_*$ and $l_{**}$.
\begin{description}
\item[\bf (1)]
$2 m_1-1\ge l_{**}\ge l_{*}\ge 1$
and $m_0\ge l_{**}$.

\item[\bf (2)]
There exists a sequence of Clifford unitaries $U({g}_0), \ldots, U({g}_{m_1})$ 
such that
$2 l_*=\rank P_{{\cal V}}^{\top} J P_{{\cal V}}$ and $l_{*}+l_{**}=\dim {\cal V}$.
\end{description}
\end{lemm}

The relation {\bf (2)}$\Rightarrow${\bf (1)} is shown as follows.
The inequality $l_{**}\ge l_{*}$ follows from 
\eqref{G02}, and
the inequality $l_{*}\ge 1$ follows from \eqref{G0}.
Since $\rank P_{{\cal V}}^{\top} J P_{{\cal V}} =2 l_*$,
there exist $l_{**}$ independent vectors 
$x_1, \ldots,x_{l_{**}} \in {\cal V}$ such that 
$\langle x_j, J x_i\rangle=0 $ for $i,j=1, \ldots l_{**}$.
Since the number of such vectors is upper bounded by $m_0$,
we have $m_0 \ge l_{**}$.
The relation $2m_1-1\ge l_{**}$ follows from \eqref{G1}.
The opposite direction {\bf (1)}$\Rightarrow${\bf (2)} 
will be shown later by using after Lemma \ref{L1}.
Lemmas \ref{L2-1}, \ref{L2-2}, and \ref{L6} imply that
the worst quantum capacity is $(m_0-2m_1 +1)\log q$, which shows Theorem \ref{Th2}.


\section{Basis-linear network model}
To construct a concrete network model to satisfy Condition {\bf (2)} given in Lemma \ref{L6},
we consider a special class of Clifford networks, called basis-linear networks.
In basis-linear networks, 
we assume that each Clifford unitary $U_i$ is characterized as
the basis exchange caused by an invertible matrix $\bar{g}_i$ on $\FF_q^{m_0}$,
which is similar to the case of CSS (Calderbank-Shor-Steane) code \cite{CS96,Steane96}.
That is, the Clifford unitary $U_i$ is given as the unitary $\bar{U} (\bar{g}_i)$
defined by
$
\bar{U} (\bar{g})|\bx \rangle
=|\bar{g} \bx \rangle$.
Its action on the Fourier basis 
$\{\ket{\by}_F\}_{\by \in \FF_q^{m_0}}$
is characterized as
$
\bar{U} (\bar{g})|\by \rangle_F
=|[\bar{g}]_F \by \rangle_F$,
where $[\bar{g}]_F$ is defined as 
the transpose $(\bar{g}^{-1})^{\top}$ of the inverse matrix $\bar{g}^{-1}$
\cite[Appendix A]{SH18-2}.
Hence, we have
\begin{align}
\bar{U} (\bar{g})=
U
\left(
\left(
\begin{array}{cc}
\bar{g}& 0 \\
0 & [\bar{g}]_F
\end{array}
\right)
\right).
\end{align}

Let $\bar{e}_i$ be the vector in $\FF_q^{m_0}$ that has only one nonzero element $1$ in the $i$-th entry.
By using the vector 
$\bar{e}_1= (1, 0, \ldots, 0) \in \FF_q^{m_0}$,
the vectors $v_1,\ldots, v_{2m_1}$ are written as
$
v_i= (\bar{v}_i, {\bf 0} ) $ and 
$v_{m_1+i}= ({\bf 0}, \bar{v}_i')$
with
$\bar{v}_i := \bar{g}_0^{-1}\cdots \bar{g}_{i-1}^{-1} \bar{e}_1 $ and
$\bar{v}_i':= \bar{g}_0^{\top}\cdots \bar{g}_{i-1}^{\top} \bar{e}_1$ 
for 
$i=1, \ldots, m_1$.
We define the matrices
$\bar{V}$ and $\bar{V}'$ as 
$(\bar{v}_1, \ldots, \bar{v}_{m_1})$
and
$(\bar{v}_1', \ldots, \bar{v}_{m_1}')$.
Then, 
we have 
\begin{align}
m_{*}\!= \!\rank (\bar{V}')^{\top} \bar{V},~ 
m_{**}\! =\! \rank \bar{V}+\rank \bar{V}'\!-\!m_{*}.\label{K2}
\end{align}

\begin{lemm}\label{L1}
The following conditions are equivalent for three integers $l_1$, $l_2$, and $l_3$.
\begin{description}
\item[\bf (1)]
$m_1\ge l_1\ge l_3\ge 1$, $m_1\ge l_2\ge l_3\ge 1$, 
and $m_0\ge l_1+l_2-l_3$.

\item[\bf (2)]
There exists a sequence of invertible matrices $\bar{g}_0, \ldots, \bar{g}_{m_1}$ over finite field $\FF_q$ such that
$\rank \bar{V}=l_1$, $\rank \bar{V}'=l_2$, and $\rank (\bar{V}')^{\top} \bar{V} =l_3$.
\end{description}
\end{lemm}
Lemma \ref{L1} is shown in Appendix E with the concrete construction of a basis-linear network to satisfy 
Condition {\bf (2)} in Lemma \ref{L1}.

When Condition {\bf (1)} of Lemma \ref{L6} holds,
Condition {\bf (1)} of Lemma \ref{L1} holds with the condition
$l_*=l_3$ and $l_{**}=l_1+l_2-l_3$.
Since 
Condition {\bf (2)} in Lemma \ref{L1} implies 
Condition {\bf (2)} of Lemma \ref{L6} with this condition,
we obtain the direction {\bf (1)}$\Rightarrow${\bf (2)} in Lemma \ref{L6}, which completes the proof of Theorem \ref{Th2}.

\section{Discussion}
We have shown that the quantum capacity is not smaller than 
$(m_0-2m_1+1)\log d$
when 
the sender has $m_0$ outgoing channels,
the receiver has $m_0$ incoming channels, 
each intermediate node applies invertible unitary,
only $m_1$ channels are corrupted in our quantum network model, 
and 
other non-corrupted channels are noiseless.
Our result holds with the following two cases.
In the first case, the unitaries on intermediate nodes are arbitrary and
the corruptions on the $m_1$ channels are individual.
In the second case, the unitaries on intermediate nodes are restricted to Clifford operations and
the corruptions on the $m_1$ channels are adaptive, i.e.,
the attacker is allowed to have a quantum memory.
Further, our code in the second case
realizes the noiseless communication even with the single-shot setting,
and depends only on the node operations, the network topology, and the places of the $m_1$ corrupted channels.
That is, it is independent of Eve's operation on the $m_1$ corrupted channels.
This code utilizes the following structure of this model.
The error in the first corrupted channel
can be concentrated to one quantum system.
However, the errors of the computation basis and 
the Fourier basis in another corrupted channel
split to two quantum systems in general.
Hence, $2m_1-1$ quantum systems are corrupted in the worst case.
As explained in Appendix,
the first case has been shown by the analysis of 
the coherent information, and  
symplectic structure including 
symplectic diagonalization on the discrete system
plays a key role in the second case.
It is an interesting remaining problem to derive the quantum capacity
when 
the operations on intermediate nodes are arbitrary unitaries
and 
the corruptions on the $m_1$ channels are adaptive.

\acknowledgments
MH was supported in part by
 Japan Society for the Promotion of Science (JSPS) Grant-in-Aid for Scientific Research (A) No. 17H01280, (B) No. 16KT0017, and Kayamori Foundation of Informational Science Advancement.

\appendix
\section*{Appendix}
The appendix is organized as follows. 
Appendix~\ref{CLA} discusses the classical case.
Appendix~\ref{PL3-1} proves Lemma \ref{L3-1}.
Appendix~\ref{AL2-2} proves Lemma \ref{L2-2}.
Appendix~\ref{AS7} gives 
the choice of vectors 
$w_1, \ldots, w_{m_0}$ and $w_1', \ldots, w_{m_0}'$
that are used in code to achieve the rate
$C \ge (m_0-m_{**})\log q$.
Since it employs symplectic diagonalization,
Appendix~\ref{AS1} summarizes a fundamental knowledge of 
symplectic diagonalization.
Appendix~\ref{AL1} proves Lemma \ref{L1}.

\section{Classical network model}\label{CLA}
We consider a classical network model as follows.
Every channel transmits a system whose number of elements is $d$
by one use of the network.
The sender has $m_0$ outgoing channels.
Node operations are not necessarily linear but are invertible.
$m_1$ channels are corrupted.
Hence,
we can assume that $m_1$ corruptions are done sequentially.
Let $X_{i-1}'$ and $X_i$ be the whole information before and after the $i$-th corruption, respectively.  
{Then,} $X_i'$ is written as $f_i (X_{i})$ by using an invertible function $f_i$.
Here, we denote the input and output information of this network by 
{$X_0$ and $X_{m_1}'$},
respectively.  
{Since the channel capacity of classical communication is given by the maximum mutual information between the input information and the output information,}
it is sufficient to show that
\begin{align}
I({X_0;X_{m_1}'})= I({X_0;X_{m_1}}) \ge (m_0-m_1)\log d \label{F1}
\end{align}
with a certain distribution $P_{{X}_0}$ of ${X}_0$.

Now, we set the distribution $P_{X_0}$ of $X_0$ to be the uniform distribution,
which implies that
\begin{align}
H(X_0) =m_0 \log d.  \label{F2}
\end{align}
From the network structure, we find the  relation 
$H(X_i|X_{i+1}) \le \log d$.
The chain rule of conditional entropy implies that
\begin{align}
& H({X_0}|X_{m_1}) 
\le
H(X_0\ldots X_{m_1-1}| X_{m_1})\nonumber \\
=&\sum_{l=0}^{m_1-1}H(X_{l}|X_{l+1} X_{l+2} \ldots X_{m_1})\nonumber \\
\le& {\sum_{l=0}^{m_1-1}H(X_{l}|X_{l+1})}
\le m_1 \log d.\label{F3}
\end{align}
The combination of \eqref{F2} and \eqref{F3} 
yields \eqref{F1}.

\section{Proof of Lemma \ref{L3-1}}\label{PL3-1}
It is sufficient to show the case when $U_0$ is the identity matrix.
Let $\rho_{\mathrm{mix},m_0-1}$ be the completely mixed state on $\cH^{\otimes m_0-1}$.
We set the initial state to be $|0\rangle \langle 0| \otimes \rho_{\mathrm{mix},m_0-1}$.

\begin{figure}[htbp]
\includegraphics[width=8cm]{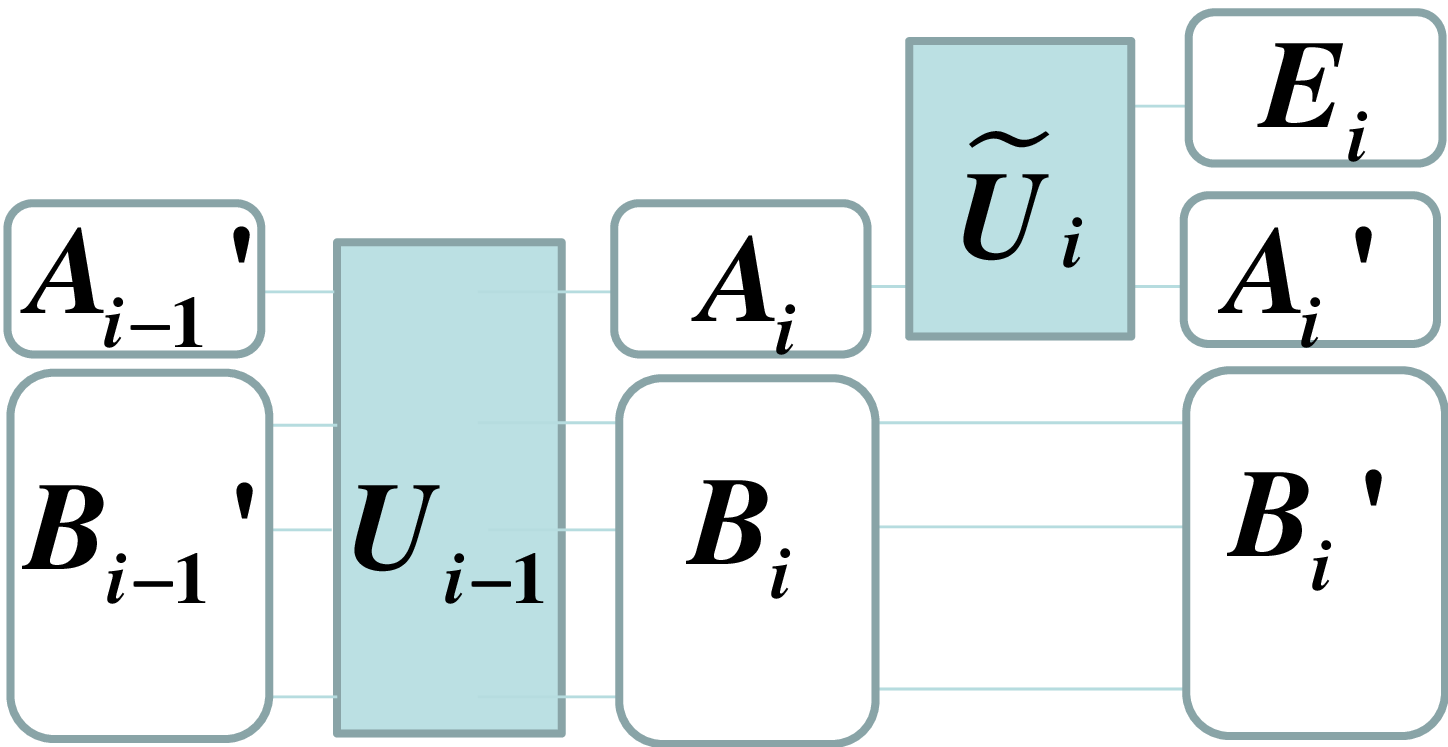}	
\caption{
Relation among systems
$A_{i-1}',B_{i-1}',$ $A_{i},B_{i},$ $A_{i}',B_{i}'$, and $E_i$.
After the application of the unitary $U_{i-1}$,
the operation $\tilde{U}_i$ is applied.}
\label{Fig3}
\end{figure}

Consider the time after the unitary $U_{i-1}$ is applied
but $\Gamma_{i}$ is not applied yet.
At this time, we denote the system to be attacked and the remaining system
by $A_i$ and $B_i$, respectively.
After the application of $\Gamma_{i}$,
we denote the system to be attacked and the remaining system
by $A_i'$ and $B_i'$, respectively.
We consider Steinspring representation $\tilde{U}_i$ of $\Gamma_i$, in which 
the output of the environment is $E_i$.
Fig. \ref{Fig3} summarizes the relation among the systems
$A_{i-1}',B_{i-1}',$ $A_{i},B_{i},$ $A_{i}',B_{i}'$, and $E_i$.

Since the state on $A_1'E_1$ is pure, we have $H(A_1')=H(E_1)$.
Hence, we have
\begin{align}
& H(A_{1}'B_1')- H(E_1)
=
H(A_{1}')+H(B_1')- H(E_1) \nonumber \\
=& H(B_1')= (m_0-1)\log d.\label{LO1}
\end{align}
As shown later, for $i=2, \ldots, m_1$,
we have
\begin{align}
H(A_{i}'B_i')- H(E_i)
\ge
H(A_{i-1}'B_{i-1}')- 2 \log  d.\label{LO2}
\end{align}
Combining \eqref{LO1} and \eqref{LO2}, we have
\begin{align}
& I_c(|0\rangle \langle 0| \otimes \rho_{\mathrm{mix},m_0-1}, \Lambda(U_0,\Gamma_1, U_1,\ldots, \Gamma_{m_1},U_{m_1}) ) \nonumber \\
=&
H(A_{m_1}'B_{m_1}')- H(E_1, \ldots, E_{m_1}) \nonumber \\
\ge &
(m_0-2m_1+1)\log d.
\end{align}

Now, we show \eqref{LO2}.
Consider the purification of $\rho_{A_i'E_i}$ by using the reference system $R$.
Then, $H(R)=H(A_i'E_i)=H(A_i)$.
Since $H(A_i)+H(A_i')-H(E_i)
=H(R)+H(A_i')-H(R A_i')
=I(R;A_i')\ge 0 $,
we have 
\begin{align}
H(E_i)\le H(A_i)+ H(A_i').\label{N1}
\end{align}
Thus, 
\begin{align}
& H(A_{i}'B_i')- H(E_i)
\stackrel{(a)}{\ge}
H(A_{i}'B_i')- H(A_i)- H(A_i')\nonumber \\
=&
H(B_{i}')+H(A_i')- H(A_i)- H(A_i')
-I(A_{i}'; B_i') \nonumber \\
=&
H(A_{i}' B_{i}' E_i)- H(A_{i}' E_i)
+I(A_{i}'E_i;B_{i}')\nonumber  \\
&+H(A_i')- H(A_i)- H(A_i')
-I(A_{i}'; B_i') \nonumber \\
\stackrel{(b)}{=}&
H(A_{i} B_{i} )- H(A_{i})
+I(E_i;B_{i}'|A_{i}')
- H(A_i) \nonumber \\
\stackrel{(c)}{\ge}&
H(A_{i}B_{i})- 2 H(A_i)
=
H(A_{i-1}'B_{i-1}')- 2 H(A_i) \nonumber \\
\ge & H(A_{i-1}'B_{i-1}')- 2 \log  d.
\end{align}
Here,
$(a)$ follows from \eqref{N1}, 
$(b)$ follows from $H(A_{i}' B_{i}' E_i)=H(A_{i} B_{i} )$
and $H(A_{i}' E_i)= H(A_{i})$,
and
$(c)$ follows from $I(E_i;B_{i}'|A_{i}')\ge 0$.

\QEDA

\section{Proof of Lemma \ref{L2-2}}\label{AL2-2}
To show Lemma \ref{L2-2}, we prepare the following lemma, which will be shown in the end of this Appendix.
\begin{lemm}\label{L6B}
When a channel $\Lambda_A$ is entanglement-breaking, 
a channel $\Lambda_B$ satisfies the condition
\begin{align}
\max_{\rho}I_c(\rho, \Lambda_A\otimes \Lambda_B)
=
\max_{\tau}I_c(\tau, \Lambda_B).
\end{align}
\end{lemm}

For the preparation of Lemma \ref{L2-2},
we define the notation for the generalized Pauli channel as follows.
Given a distribution $P$ on $\FF_q^{2l}$, we define the channel $\Lambda_P$
as
\begin{align}
\Lambda_P(\rho):=
\sum_{(\bs,\bt)\in  \FF_q^{2l}}
P(\bs,\bt) \bW(\bs,\bt) \rho \bW(\bs,\bt)^{-1}.
\end{align}

Then, we denote 
the uniform distribution over $\FF_q^{2 m}$ and
the uniform distribution over the subset $\{(\bs,0)\}_{\bs \in \FF_q^{m}}$
by $P_{\mathrm{mix},m}$ and $P_{\mathrm{mix} Z,m}$, 
respectively.
Hence,
the channels on the attacked edges are the Pauli channel $\Lambda_{P_{\mathrm{mix},1}}$.

We assume that the sender applies the encoding unitary $U_e$ before transmission
and the receiver applies the decoding unitary $U_d$ after the reception.
Considering the output behaviors on the computation basis and on the Fourier basis,
we find that 
the channel from the sender to the receiver is given as 
$\mathrm{Id}_{m_0-m_{**}}
\otimes
\Lambda_{P_{\mathrm{mix},m_{*}}} \otimes
\Lambda_{P_{\mathrm{mix} Z,m_{**}-m_{*}}} $.
Since $P_{\mathrm{mix} Z,m}$ is the pinching channel with respect to the measurement on the Fourier basis,
we find that the channel 
$\Lambda_{P_{\mathrm{mix},m_{*}}} \otimes \Lambda_{P_{\mathrm{mix} Z,m_{**}-m_{*}}} $ 
is entanglement-breaking.


Since the channel 
$\Lambda_{P_{\mathrm{mix},m_{*}}} \otimes \Lambda_{P_{\mathrm{mix} Z,m_{**}-m_{*}}} $ 
 is entanglement-breaking,
Lemma \ref{L6B} guarantees that
\begin{align}
& \max_{\rho} I_c(\rho, 
(\mathrm{Id}_{m_0-m_{**}}
\otimes
\Lambda_{P_{\mathrm{mix},m_{*}}} \otimes
\Lambda_{P_{\mathrm{mix} Z,m_{**}-m_{*}}} )^{\otimes n}
 )\nonumber \\
 =&
\max_{\rho' }
I_c(\rho',\mathrm{Id}_{m_0-m_{**}}^{\otimes n})
=n (m_0-m_{**}) \log q ,
\end{align}
where $I_c(\rho,\Lambda)$ is the coherent information.

Since the maximum transmission rate is upper bounded by the maximum coherent information, 
we obtain the converse part. 

\QEDA


\noindent{\it Proof of Lemma \ref{L6B}:\quad}
Let $A$ and $B$ ($A'$ and $B'$) be the input (output) systems of 
$\Lambda_A$ and 
$\Lambda_B$, respectively.
We choose a state $\rho_{AB}$ on $AB$.
Let $C$ be the reference system of the state $\rho_{AB}$ so that
$\rho_{ABC}$ is the purification of $\rho_{AB}$.
Let $\rho'$ be the output system on the whole system of $A'$, $B'$ and $C$.

Since $\Lambda_A$ is entanglement-breaking,
it is written as 
$ \Lambda_A(\sigma)= \sum_{a} \rho_{A'=a} \Tr M_a \sigma$,
where $\{M_a\}$ is a POVM and $\rank M_a=1$.
Hence, 
 $\Lambda_A( \rho_{ABC})$ is written as 
 $\sum_{a} P_{A'=a} \rho_{BC|A'=a} \otimes \rho_{A'=a}$,
 where $P_{A'=a}:= \Tr M_a \rho_{A}$ and 
 $\rho_{BC|A'=a}:= \Tr_A M_a \rho_{ABC}/P_{A'=a}$.
 Hence, $\rank \rho_{BC|A'=a}=1$.
 
Then, we denote $\Lambda_B(\rho_{BC|A'=a}) $ by $\rho_{B'C|A'=a}'$.
The coherent information 
$I_c( \rho_{AB}, \Lambda_A\otimes \Lambda_B)$
equals 
$D(\rho_{A'B'C}'\| \rho_{A'B'}'\otimes I_C )$,
which is evaluated as
\begin{align}
&D(\rho_{A'B'C}'\| \rho_{A'B'}'\otimes I_C )\nonumber \\
=&
D\Bigg(\sum_{a} P_{A'=a} \rho_{B'C|A'=a}' \otimes \rho_{A'=a}\Bigg\| \nonumber \\&
\qquad\sum_{a} P_{A'=a} \rho_{B'|A'=a}' \otimes \rho_{A'=a}\otimes I_C\Bigg) \nonumber \\
\le &
D\Bigg(\sum_{a} P_{A'=a} \rho_{B'C|A'=a}' \otimes |a\rangle \langle a| \Bigg\| \nonumber \\&
\qquad\sum_{a} P_{A'=a} \rho_{B'|A'=a}' \otimes  |a\rangle \langle a|  \otimes I_C\Bigg) \nonumber \\
= &
\sum_{a} P_{A'=a} D( \rho_{B'C|A'=a}' \|\rho_{B'|A'=a}'  \otimes I_C).
\end{align}
The inequality follows from the information processing inequality for the map 
$  |a\rangle \langle a|  \mapsto \rho_{A'=a}$. 

Since $\rank \rho_{BC|A'=a}=1$, 
$\rho_{BC|A'=a}$ is a purification of $\rho_{B|A'=a}$.
$D( \rho_{B'C|A'=a}' \|\rho_{B'|A'=a}'  \otimes I_C) $ equals the coherent information
$I_c( \rho_{B|A'=a}, \Lambda_B) $.
Hence, we have
\begin{align}
&I_c( \rho_{AB}, \Lambda_A\otimes \Lambda_B)\nonumber \\
\le &
\sum_{a} P_{A'=a} I_c( \rho_{B|A'=a}, \Lambda_B),
\end{align}
which implies that 
\begin{align}
\max_{\rho}I_c(\rho, \Lambda_A\otimes \Lambda_B)
\le
\max_{\tau}I_c(\tau, \Lambda_B).
\label{ineq:direct}
\end{align}

Next, we show the converse inequality 
\begin{align}
\max_{\rho}I_c(\rho, \Lambda_A\otimes \Lambda_B)
\ge
\max_{\tau}I_c(\tau, \Lambda_B).
\label{ineq:converse}
\end{align}
For any state $\tau$ on the system $B$, 
define $\rho_{AB} = \rho_A \otimes \tau$ where $\rho_A$ is a pure state.
Then, we have
\begin{align}
I_c(\rho_{AB} , \Lambda_A\otimes \Lambda_B)
=
I_c(\tau, \Lambda_B).
\end{align}
Therefore, we obtain \eqref{ineq:converse}.
\QEDA

\section{Choice of vectors 
$w_1, \ldots, w_{m_0}$ and $w_1', \ldots, w_{m_0}'$}\label{AS7}
\subsection{Construction except for $w_{m_0-m_{**}+m_{*}+1}', \ldots, w_{m_0}'$}
Now, we choose 
we can choose $2 m_0$ independent vectors 
$w_1, \ldots, w_{m_0}$ and $w_1', \ldots, w_{m_0}'$ satisfying the following conditions.
(i) 
$\langle w_i',J w_j \rangle =\delta_{i,j}$ and
$\langle w_i,J w_j \rangle =\langle w_i',J w_j' \rangle =0$ 
for $i,j=1, \ldots, m_0$.
(ii)
The space $ {\cal V}+J{\cal V}$ is spanned by 
$w_{m_0-m_{**}}, \ldots, w_{m_0}$ and $w_{m_0-m_{**}}', \ldots, w_{m_0}'$.

Define ${\cal V}_2:=\Ker P_{{\cal V}}^{\top} J P_{{\cal V}}\cap {\cal V}$.
We choose another subspace ${\cal V}_1$ of ${\cal V}$ such that
${\cal V}_1\oplus {\cal V}_2={\cal V}$.
Since $P_{{\cal V}}^{\top} J P_{{\cal V}}$ is non-degenerate on ${\cal V}_1$,
we can choose 
$m_{*}$ independent vectors $w_{m_0-m_{**}+1}, \ldots, 
w_{m_0-m_{**}+m_{*}} \in 
{\cal V}_1\subset \FF_q^{2m_0}$ and
other $m_{*}$ independent vectors $w_{m_0-m_{**}+1}', \ldots, 
w_{m_0-m_{**}+m_{*}}' \in 
{\cal V}_1\subset \FF_q^{2m_0}$ such that
$\langle w_{i}', J w_j \rangle =\delta_{i,j}$,
$\langle w_i,J w_j \rangle =\langle w_i',J w_j' \rangle =0$
for 
$i,j= m_0-m_{**}+1, \ldots, m_0-m_{**}+m_{*}$.

Let $w_{m_0-m_{**}+m_*+1}, ..., w_{m_0}$ be a basis of ${\cal V}_2$.
We define 
${\cal V}_3:=\Ker P_{{\cal V}}^{\top} J P_{{\cal V}}$.
We have the direct sum ${\cal V}_1\oplus {\cal V}_3=\FF_q^{2m_0} $.
Based on them, as shown in Subsection \ref{AS6}, we can choose
independent vectors 
$w_{m_0-m_{**}+m_{*}+1}', \ldots, w_{m_0}'
\in {\cal V}_3\subset \FF_q^{2m_0}$
to satisfy the conditions
$\langle w_i', J w_j \rangle =\delta_{i,j}$ and $\langle w_i', J w_j' \rangle =0$
for $i,j= m_0-m_{**}+m_{*}+1, \ldots, m_0$.

The subspace 
${\cal V}_4$ is defined as the space spanned by 
$w_{m_0-m_{**}+1}, \ldots, w_{m_0}$ and $w_{m_0-m_{**}+1}', \ldots, w_{m_0}'$.
Choosing a projection $P_{{\cal V}_4}$ onto ${\cal V}_4$,
we define the subspace ${\cal V}_5:=\Ker P_{{\cal V}_4}^{\top} J$.
Since the dimension of the image of $P_{{\cal V}_4}^{\top} J$ is $2m_{**}$,
that of ${\cal V}_5$ is $2(m_0-m_{**})$.
Also, there is no cross term in $J$ between 
${\cal V}_4$ and ${\cal V}_5$
because $\langle v,J v'\rangle
= \langle P_{{\cal V}_4}v,J v'\rangle 
= \langle v,P_{{\cal V}_4}^{\top} J v'\rangle =0$
for $v \in {\cal V}_4,v'\in {\cal V}_5$.
Thus, the rank of $P_{{\cal V}_5}^{\top} J P_{{\cal V}_5} $ is $2(m_0-m_{**})$,
where $P_{{\cal V}_5}$ is a projection to ${\cal V}_5$.
Hence, 
we choose $2(m_0-m_{**})$ independent vectors 
$w_1, \ldots, w_{m_0-m_{**}} \in {\cal V}_5$
and
$w_1', \ldots, w_{m_0-m_{**}}' \in {\cal V}_5$
such that 
$\langle w_i',J w_j \rangle =\delta_{i,j}$ 
and
$\langle w_i,J w_j \rangle =\langle w_i',J w_j' \rangle =0$ 
for $i,j=1, \ldots, m_0-m_{**}$.

Since there is no cross term in $J$ between 
${\cal V}_4$ and ${\cal V}_5$,
the chosen vector 
$w_1, \ldots, w_{m_0} \in \FF_q^{2m_0}$
and
$w_1', \ldots, w_{m_0}' \in\FF_q^{2m_0}$ satisfy the conditions
$\langle w_i',J w_j \rangle =\delta_{i,j}$ 
and
$\langle w_i,J w_j \rangle =\langle w_i',J w_j' \rangle =0$ 
for $i,j=1, \ldots, m_0$.
Therefore, we can choose a symplectic matrix $g_*$ such that
$g_* e_i=  w_i$ and $g_* e_{m_0+i} =  w_i'$ for $i=1, \ldots , m_0$
because the vectors $w_i$ and $w_i'$ with $i=1, \ldots , m_0$
have the same symplectic structure as the vectors $e_1,\ldots, e_{2m_0}$. 

\subsection{Symplectic diagonalization}\label{AS1}
For the choice of $w_{m_0-m_{**}+m_{*}+1}', \ldots, w_{m_0}'$,
we prepare fundamental knowledge for symplectic diagonalization
in the finite dimensional system.
Assume that ${\cal V}$ is a finite-dimensional vector space over a finite field $\FF_q$.
We consider a bilinear form $Q$ from  
${\cal V}\times {\cal V}$ to $\FF_q$.
Given an element $v \in V$,
$Q(v,\cdot)$ can be regarded as an element of the dual space 
${\cal V}^*$ of ${\cal V}$.
In this sense, $Q$ can be regarded as a linear map from 
${\cal V}$ to ${\cal V}^*$.
A bilinear form $Q$ from 
${\cal V}\times {\cal V}$ to $\FF_q$ is called 
anti-symmetric when
$Q(v_1,v_2)=-Q(v_2,v_1)$ and 
$Q(v_1,v_1)=0$ for $v_1,v_2 \in {\cal V}$.

\begin{lemm}\label{LC-1}
Assume that an anti-symmetric bilinear form
$Q$ is surjective, i.e., $\Ker Q$ is $\{0\}$.
Then, the dimension of ${\cal V}$ is an even number $2k$.
There exists a basis $w_1,\ldots, w_{k},w_1',\ldots, w_{k}' \in {\cal V} $ such that
$Q(w_i',w_j)=\delta_{i,j}$ and
$Q(w_i,w_j)=Q(w_i',w_j')=0$ for
$i,j=1,\ldots,k$.
\end{lemm}

\noindent{\it Proof of Lemma \ref{LC-1}:\quad}
Such a basis can be chosen inductively.
We choose a non-zero vector $w_1 \in {\cal V}$.
Since $Q$ is surjective,
we can choose another non-zero vector $w_1' \in {\cal V}$
such that $Q(w_1',w_1)=1$.

Due to the assumption of induction,
we have vectors $w_1,\ldots, w_{l},w_1',\ldots, w_{l}' \in {\cal V} $ such that
$Q(w_i',w_j)=\delta_{i,j}$ and
$Q(w_i,w_j)=Q(w_i',w_j')=0$ for
$i,j=1,\ldots,l$.
Then, we define the subspace ${\cal V}_l:=
\{v \mid
Q(w_1,v)=\cdots =Q(w_{l},v)=Q(w_1',v)=\cdots = Q(w_{l}',v)=0\}$.
Also, we define
the subspace ${\cal V}_l'$ spanned by 
$w_1,\ldots, w_{l},w_1',\ldots, w_{l}' \in {\cal V} $.
Since ${\cal V}_l\cap {\cal V}_l' = \{0\}$,
we have the direct sum ${\cal V}={\cal V}_l\oplus {\cal V}_l'$.
We choose a non-zero vector $w_{l+1} \in {\cal V}_l$.
Since $Q$ is surjective,
we can choose another non-zero vector $v_{l+1} \in {\cal V}$
such that $Q(v_{l+1},w_{l+1})=1$.
Also, we have $Q(w_{i},w_{l+1})=Q(w_{i}',w_{l+1})=0$ for 
$i=1,\ldots,l$.
Based on the direct sum ${\cal V}= {\cal V}_l \oplus {\cal V}_l'$,
we have the decomposition $ v_{l+1}=v_{l+1}'+w_{l+1}'$ with
$v_{l+1}'\in {\cal V}_l'$ and $w_{l+1}'\in {\cal V}_l$.
Since $w_{l+1}'\in {\cal V}_l$, 
we have $Q(w_{i},w_{l+1}')=Q(w_{i}',w_{l+1}')=0$ for 
$i=1,\ldots,l$.
Since $Q(v_{l+1}',w_{l+1})=0$, we have
$Q(w_{l+1}',w_{l+1})=Q(v_{l+1},w_{l+1})=1$.
Therefore, we obtain a desired basis.
\QEDA

\subsection{Choice of $w_{m_0-m_{**}+m_{*}+1}', \ldots, w_{m_0}'$}\label{AS6}
Using symplectic diagonalization,
we show that we can choose
independent vectors 
$w_{m_0-m_{**}+m_{*}+1}', \ldots, w_{m_0}'
\in {\cal V}_3\subset \FF_q^{2m_0}$,
inductively.
First, we choose $u_{m_0-m_{**}+m_{*}+1}$
to satisfy
the condition $\langle u_{m_0-m_{**}+m_{*}+1}, J w_j \rangle =\delta_{m_0-m_{**}+m_{*}+1,j}$
for 
$j= m_0-m_{**}+m_{*}+1, \ldots, m_0$.
Based on the direct sum 
${\cal V}_1\oplus {\cal V}_3 = \FF_q^{2m_0}$, we decompose   
$u_{m_0-m_{**}+m_{*}+1}= u_{m_0-m_{**}+m_{*}+1}'+w_{m_0-m_{**}+m_{*}+1}'$ such that $u_{m_0-m_{**}+m_{*}+1}' \in {\cal V}_1, 
w_{m_0-m_{**}+m_{*}+1}'\in  {\cal V}_3$.
Since $\langle u_{m_0-m_{**}+m_{*}+1}', J w_{j} \rangle =0$,
we have $\langle w_{m_0-m_{**}+m_{*}+1}', J w_j \rangle 
=\delta_{m_0-m_{**}+m_{*}+1,j}$
for $j= m_0-m_{**}+m_{*}+1, \ldots, m_0$.

Next, from vectors 
$w_{m_0-m_{**}+m_{*}+1}', \ldots, w_{l}'
\in {\cal V}_3\subset \FF_q^{2m_0}$,
we choose $u_{l+1}$
to satisfy
the conditions 
\begin{align}
\langle u_{l+1}, J w_j \rangle =\delta_{i,j},
~\langle u_{l+1}, J w_i' \rangle =0
\label{Con1}
\end{align}
for 
$j= m_0-m_{**}+m_{*}+1, \ldots, m_0$
and
$i= m_0-m_{**}+m_{*}+1, \ldots, l$.
Based on the direct sum
${\cal V}_1\oplus {\cal V}_3 = \FF_q^{2m_0}$, we decompose   
$u_{l+1}= u_{l+1}'+\bar{w}_{l+1}'$ such that $u_{l+1}' \in {\cal V}_1, 
\bar{w}_{l+1}'\in  {\cal V}_3$.
Then, we define 
$w_{l+1}':=\bar{w}_{l+1}'+ \sum_{i=m_0-m_{**}+m_{*}+1}^l 
\langle u_{l+1}', J w_{i}' \rangle w_{i}$.
Hence, we have
$u_{l+1}= u_{l+1}'- (\sum_{i=m_0-m_{**}+m_{*}+1}^l 
\langle u_{l+1}', J w_{i}' \rangle w_{i})+w_{l+1}'$.
Since $u_{l+1}' - \sum_{i=m_0-m_{**}+m_{*}+1}^l 
\langle u_{l+1}', J w_{i}' \rangle w_{i} \in {\cal V}$
and $w_{j} \in  {\cal V}_2=\Ker P_{{\cal V}}^{\top} J P_{{\cal V}}\cap {\cal V}$
for $j= m_0-m_{**}+m_{*}+1, \ldots, m_0$,
we have the relation
\begin{align}
& \langle u_{l+1}'- \sum_{i=m_0-m_{**}+m_{*}+1}^l 
\langle u_{l+1}', J w_{i}' \rangle w_{i}, J w_{j} \rangle 
\nonumber \\
=&
\langle P_{{\cal V}} (u_{l+1}'- \sum_{i=m_0-m_{**}+m_{*}+1}^l 
\langle u_{l+1}', J w_{i}' \rangle w_{i}), J P_{{\cal V}} w_{j} \rangle
\nonumber \\
 =&
\langle u_{l+1}'- \sum_{i=m_0-m_{**}+m_{*}+1}^l 
\langle u_{l+1}', J w_{i}'\rangle w_i,P_{{\cal V}}^{\top} J P_{{\cal V}} w_{j} \rangle =
0.
\label{Con2}
\end{align}
For $i'= m_0-m_{**}+m_{*}+1, \ldots, l$,
we have
\begin{align}
&\langle u_{l+1}'- \sum_{i=m_0-m_{**}+m_{*}+1}^l 
\langle u_{l+1}', J w_{i}'\rangle w_i, J w_{i'}' \rangle
\nonumber \\
=&\langle u_{l+1}',J w_{i'}' \rangle
- \sum_{i=m_0-m_{**}+m_{*}+1}^l 
\langle u_{l+1}', J w_{i}'\rangle \langle w_{i}, J w_{i'}' \rangle
\nonumber \\
=& \langle u_{l+1}',J w_{i'}' \rangle
-\langle u_{l+1}',J w_{i'}' \rangle
= 0.
\label{Con3}
\end{align}
The combination of \eqref{Con1}, \eqref{Con2}, and \eqref{Con3},
implies the relations
$\langle w_{l+1}', J w_j \rangle =\delta_{l+1,j}$
and $\langle w_{l+1}', J w_{i'}' \rangle =0$.

Notice that $\langle w_{l+1}', J w_{l+1}' \rangle =0$
holds since $J$ is anti-symmetric.

\section{Proof of Lemma \ref{L1}}\label{AL1}
Now, we show {\bf (2)}$\Rightarrow${\bf (1)}.
Since the relations 
$m_1\ge l_1\ge l_3$, $m_1\ge l_2\ge l_3$
are trivial, it is sufficient to show $m_0\ge l_1+l_2-l_3$ and $l_3 \ge 1$.
Since the matrix $(\bar{V}')^{\top} \bar{V}$ is not zero, we have $l_3 \ge 1$.
The relation $m_0\ge l_1+l_2-l_3$ follows from \eqref{G1} and \eqref{K2}.
Thus, we obtain the relation {\bf (2)}$\Rightarrow${\bf (1)}.

Now, we show {\bf (1)}$\Rightarrow${\bf (2)}.
We assume that $l_1 \ge l_2$.
Otherwise, we can exchange the computation basis and the Fourier basis.
We choose $\bar{g}_{m_1}$ to be the identity matrix.
For $i=1, \ldots, m_1$, we choose 
$\bar{g}_{i-1}$ to be $A_{i-1} A_i^{-1}$,
where the matrix $A_i$ is defined as follows.

$A_0$ is the identity matrix.
For $i=1,\ldots, l_3$, 
we define $A_i$
as the transposition between the 1-st entry and the $i$-th entry.
For $i=l_1+1, \ldots, m_1$,
we define $A_i$ as the identity matrix.

For $i=1,\ldots, l_2-l_3$,
we define $A_{l_3+i}$ in the following way.
For the 1-st, $l_3+2i-1$-th, and $l_3+2i$-th entries, 
it is defined as
$\left(
\begin{array}{ccc}
1 & 0 & 1 \\
1 & 1 & 0 \\
0 & 0 & 1 
\end{array}
\right)$.
For other indices $j,j'$, 
the matrix component $(A_{l_3+i})_{j,j'}$
is defined as $\delta_{j,j'}$.

For $i=1,\ldots, l_1-l_2$,
we define $A_{l_2+i}$ in the following way.
For the 1-st and $2l_2-l_3+i$-th entries, 
it is defined as
$\left(
\begin{array}{cc}
1 & 0 \\
1 & 1 
\end{array}
\right)$.
For other indices $j,j'$, 
the matrix component $(A_{l_2+i})_{j,j'}$
is defined as $\delta_{j,j'}$.

Then, for $i=1, \ldots, l_3$, we have $\bar{v}_i=\bar{e}_i$.
For $i=1, \ldots, l_2-l_3$, we have $\bar{v}_{l_3+i}=\bar{e}_1+\bar{e}_{l_3+2i-1}$.
For $i=1, \ldots, l_1-l_2$, we have $\bar{v}_{l_2+i}=\bar{e}_1+\bar{e}_{2l_2-l_3+i}$.
For $i=l_1+1, \ldots, m_1$, we have $\bar{v}_i=\bar{e}_1$.

The matrix $[A_{l_3+i}]_F$ is characterized for $i=1, \ldots, l_2-l_3$ as follows.
For the 1-st, $l_3+2i-1$-th, and $l_3+2i$-th entries, 
it is given as
$\left(
\begin{array}{ccc}
1 & -1 & 0 \\
0 & 1 & 0 \\
-1 & 0 & 1 
\end{array}
\right)$.
For other indices $j,j'$, 
$([A_{l_3+i}]_F)_{j,j'}$
is given as $\delta_{j,j'}$.

The matrix $[A_{l_2+i}]_F$ is characterized for $i=1, \ldots, l_1-l_2$ as follows.
For the 1-st and $2l_2-l_3+i$-th entries, 
it is given as
$\left(
\begin{array}{cc}
1 & -1 \\
0 & 1 \\
\end{array}
\right)$.
For other indices $j,j'$, 
the matrix component $([A_{l_2+i}]_F)_{j,j'}$ is given as $\delta_{j,j'}$.
Then, 
for $i=1, \ldots, l_3$, we have $\bar{v}_i'=\bar{e}_i$.
For $i=1, \ldots, l_2-l_3$, we have $\bar{v}_{l_3+i}'=\bar{e}_1+\bar{e}_{l_3+2i}$.
For $i=1, \ldots, l_1-l_2$, we have $\bar{v}_{l_2+i}'=\bar{e}_1$.
for $i=l_1+1, \ldots, m_1$, we have $\bar{v}_i'=\bar{e}_1$.
Therefore, we have
$\rank \bar{V}=l_1$ and $ \rank \bar{V}'=l_2$.

Also, when $j=2, \ldots, l_3$ or $j'=2, \ldots, l_3$,  
we have $((\bar{V}')^{\top} \bar{V})_{j,j}=\delta_{j,j'}$.
When $j=1,l_3+1,\ldots, m_1$ or $j'=1,l_3+1,\ldots, m_1$,  
we have
$((\bar{V}')^{\top} \bar{V})_{j,j}=1$.
Hence,
$ \rank (\bar{V}')^{\top} \bar{V}= l_3$.
Thus, we obtain the relation
{\bf (1)}$\Rightarrow${\bf (2)}.

\QEDA

\end{document}